\documentclass[a4paper]{jpconf}
\usepackage{graphicx}

\begin{document}

\title{Jet Activity in the Central Super Massive Black Hole of the Milky Way in 4th and 14th centuries and millennial periods of the climate change}

\author{Olga Piskounova}
\address{Research scientist, P.N.Lebedev Physics Institute of Russian Academy of Science, Leninski prosp. 53, 119991 Moscow, Russia}

\author{Irina Tamarkina}
\address{ Independent researcher, Moscow, PhD in history at Wisconsin-Madison Uni., USA, Postdoc at Hebrew Uni., Jerusalem}

\ead{piskunovaoi@lebedev.ru}

\begin{abstract}
This paper discusses the historical evidence for the light jets from the central Black Hole of Galaxy in the 4th and 14th centuries. We suggest that the apparitions of a "lightening cross" recorded in 312, 351, 1317 and 1377 years were caused by the line of two jets beamed back-to-back from the center of Milky Way (MW), which is crossing the visible projection of the Galaxy disk. All historical accounts give precise time and geographical locations, where from X-crosses was seen: in the vicinity of Rome, in Jerusalem, in Tver' (the vicinity of Moscow) and in night sky over Moscow, respectively). The positions of the Sun (or the Moon) were also mentioned in the Middle Age manuscripts. These positions coincide with the location of the MW center on the sky map in these specific places and dates, as it is shown with the help of Wolfram Astronomy visualization instruments. We suggest that the gap between jet activity events is about thousand years. The X cross in the sky was mentioned by Plato in {\it Timaeus}\cite{plato}, but he cannot be eyewitness and could only take this from the narratives of older ancient sages. The X cross pictures are multiply stamped on Roman coins. Nowadays, such an event would easy get the interpretation as a technological catastrophe in the atmosphere. It is also important that the 4th and the 14th centuries were times when the climate has become colder. The processes in Galaxy are often periodic because of orbital movements, so we should consider the thousand years periods in the climate changes as well. 
\end{abstract}

\section{Introduction}    
The jets from Super Massive Black Holes (SMBHs), which are visible in dark space, have been recorded in this century with the Hubble telescope \cite{hubble}. No jets from the central SMBH of Milky Way (MW) have seen yet, but some remnants of them , which still have some radiation activity, we know: Fermi bubbles \cite{fermibubbles} and giant x-ray chimneys \cite{chimneys}. 
The recent study of the reflections of flares near Milky Way center \cite{recentstudy} helpes us to conclude that the particle radiation a few handreds years ago  may be tens times bigger than now. 

\begin{figure}[htpb]
  \centering
  \includegraphics[scale=0.4]{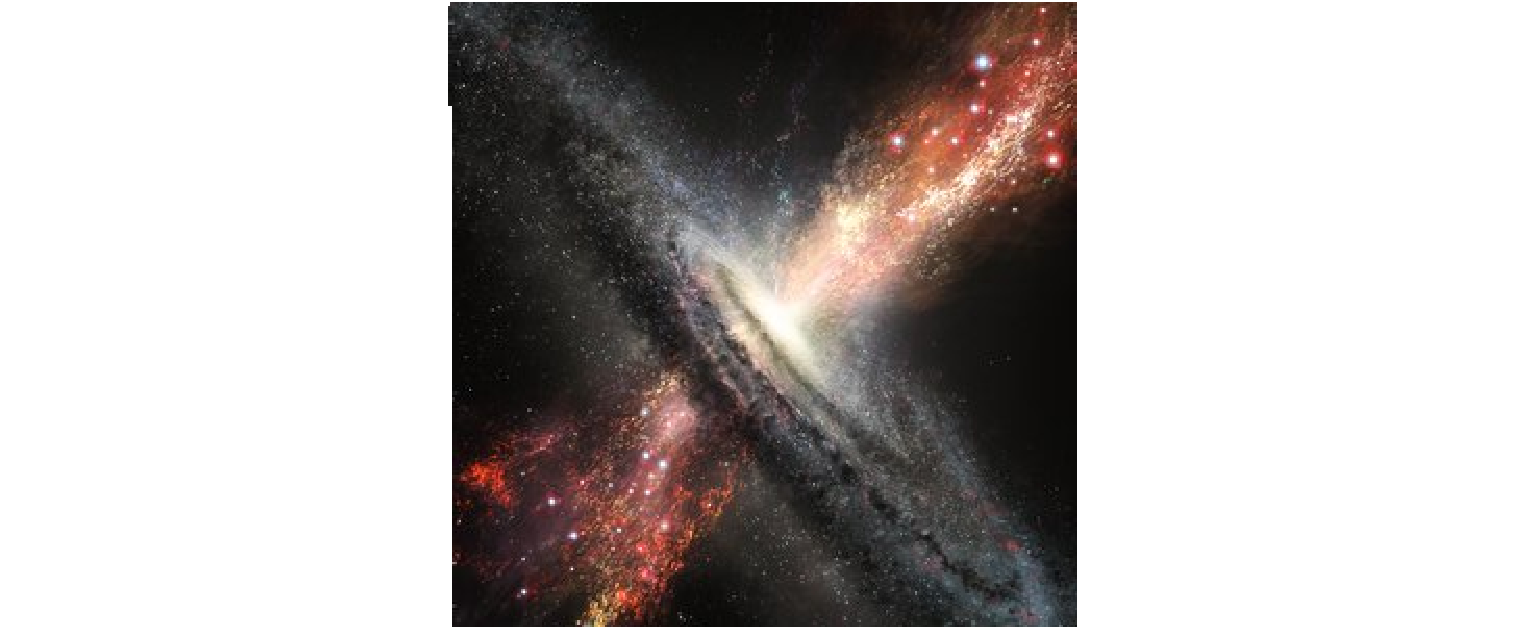}
  \caption{Artistic image of jet intersection with galaxy plane.}
  \label{artimage}
\end{figure}

Taking into account that the jets from SMBHs are suspected as main source of Dark Matter \cite{bdm}, the waves of matter can bring the periods of hot weather. There are few records in chronicles about astronomical anomalies that can be interpreted as the evidences for the jets coming from the center of Galaxy. Four events were recorded in the Middle Age manuscripts and mentioned in the later historical translations. Two intersecting lines that are forming an X-cross became visible during the times of Constantine the Great, who had been moving with his troops from Gallia to Rome in September-October of 312 CE \cite{eusebius}. Another day-time apparition of a lightening cross was described in the letter by Cyril of Jerusalem to Roman Emperor Constantius in May 351 CE \cite{cyril}. Under September of 1317 Russian chronicle that have been recording the similar day-time event with multiple light rays \cite{nikonletopis}. The forth evidence of crossing lights was found in russian chronicle under 1377 year \cite{rogozhskii}. This event took place in the night sky over Moscow three years before the Koulikovskaya battle. 

These narratives have been described the appearances of two large intersecting light lines during the day, or night time, at the clear weather. Events were rather powerful phenomena that reportedly outshone the Sun, if observed during the day. The descriptions of this astronomical anomaly most closely match the idea of the jets that beam out from the center of the MW. Since old texts give date and position of the crossing lines in the sky, it is possible to compare their location with the relation to the position of Galaxy center in the application of Wolfram Technology for Astronomy \cite{wolfram}.
  
\section{Facilities of Wolfram Astronomy and the search criteria}

\begin{figure}[htpb]
  \centering
  \includegraphics[scale=0.6]{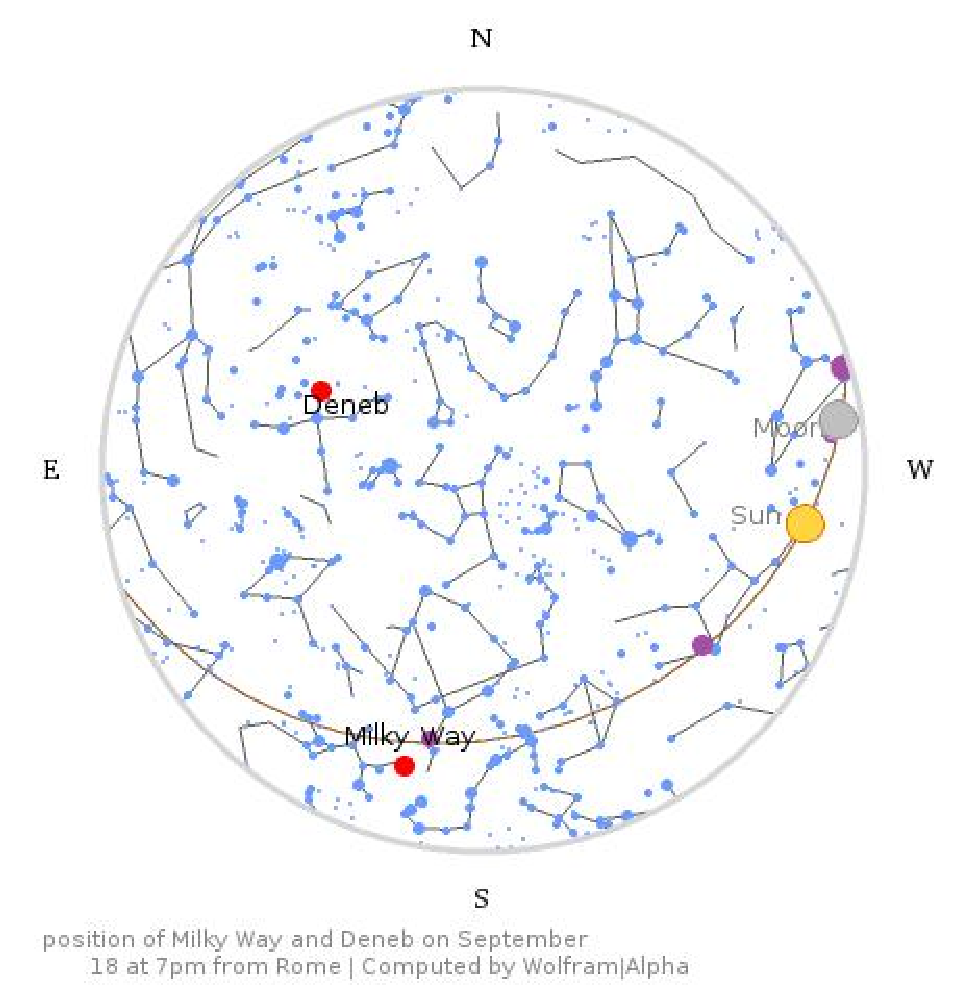}
  \caption{Position of Milky Way center and Cygnus constellation at noon on September 18.}
  \label{deneb}
\end{figure}

The Wolfram Astronomy application was examined to consider each event. Some additional tuning of application has been requested from Wolfram authors for the purposes of our research. An example of the typical Wolfram Astronomy operator looks like the following expression: "{\bf position of Milky Way at 5 am on May 7 from Jerusalem}". As we know, it is not necessary to indicate the year of event. The position of two sky objects helps us to settle the northern wing of Galaxy disk; they are MW center and Cygnus constellation, which stretches itself along the visible end of Galaxy disk. Position of Cygnus is marked with the brightest star Deneb (see figure~\ref{deneb}).

The analysis \cite{jetactivity} has to be fulfilled under the following criteria: center of the MW is above horizon and the Sun (or the Moon) position, if mentioned, should correspond to the described evidence. Unfortunately, we have the exact date of the event only for the phenomenon on May 7 of 351 CE in Jerusalem \cite{cyril}. The other events are also corresponding to the May and September periods, when Milky Way is seen in Northern hemisphere.  

\section{Analysis of events}

   \subsection{The way from Gallia to Rome}

Constantine the Great and his roman troops were marching from Arles to Rome to take battle with Maxentius, who resided in Rome. It happened between August and October of 312 CE that actually is the best time for the observation of the MW center in Europe. The center of Milky Way took place on South-East during the day time, see figure~\ref{romeevent}. The time after 3 pm seems proper for observation, when Sun is setting on the West. It is important that sun light could not interfere with light from jets. According to Constantine's biographer Eusebius cite{eusebius], who allegedly learned about the "vision" from the Emperor himself: "Once upon a time at afternoon, when the Sun turned to West, I saw with my own eyes a Sign of Cross that was built from light and lying on the Sun". 
 
\begin{figure}[htpb]
  \centering
  \includegraphics[scale=0.6]{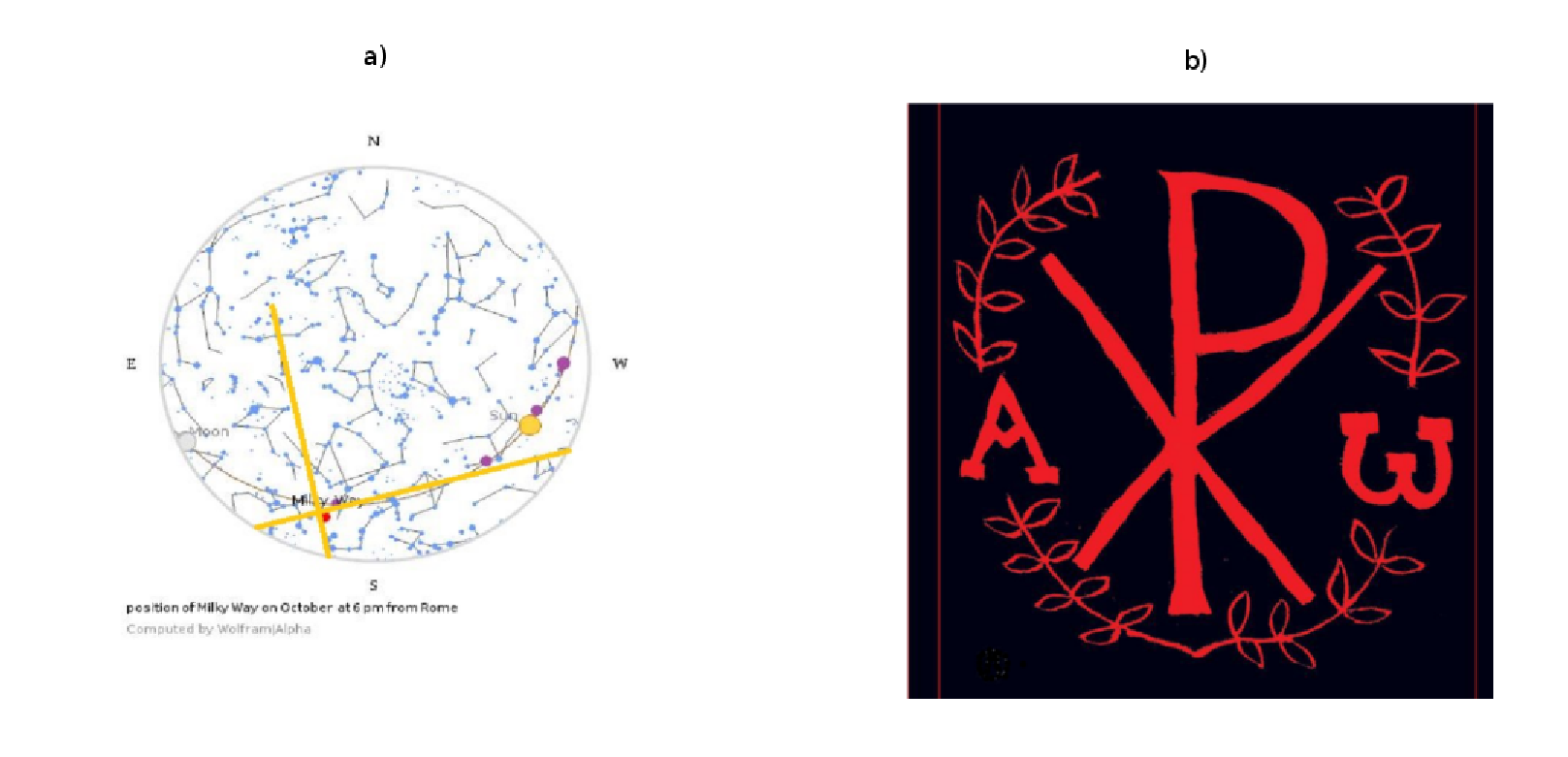}
  \caption{Position of Milky Way in October afternoon with the X cross a) and b) The Labarum symbol, where "Ro" looks like the Sun between two jets.}
 \label{romeevent}
\end{figure}

Another contemporaneous author, Lactantius, reported that Constantine has drawn afterwards this sign on the shields of his soldiers as "Chi-Rho". Finally, the vision was reflected in the Labarum - the emblem of early Christian church.


Today, the materialized witness of the vision can be seen on the Roman coins that have been minted by Constantine in 317 CE \cite{romancoins}
 
 \subsection{The letter from Jerusalem}
The next event was described in 351 CE in the letter of Cyril, bishop of Jerusalem, to Emperor Constantius (Yarnold: "In these holy days of the Easter season, on 7 May at about the third hour, a huge cross made of light appeared in the sky above holy Golgotha extending as far as the holy Mount of Olives". Since the letter to Emperor had propagandistic goal, it was dangerous just to invent this vision, because other citisen could easily disprove these news. The MW position and suggested cross on May 7 are seen in the figure~\ref{jerusalem}.
 
\begin{figure}[htpb]
  \centering
  \includegraphics[scale=0.6]{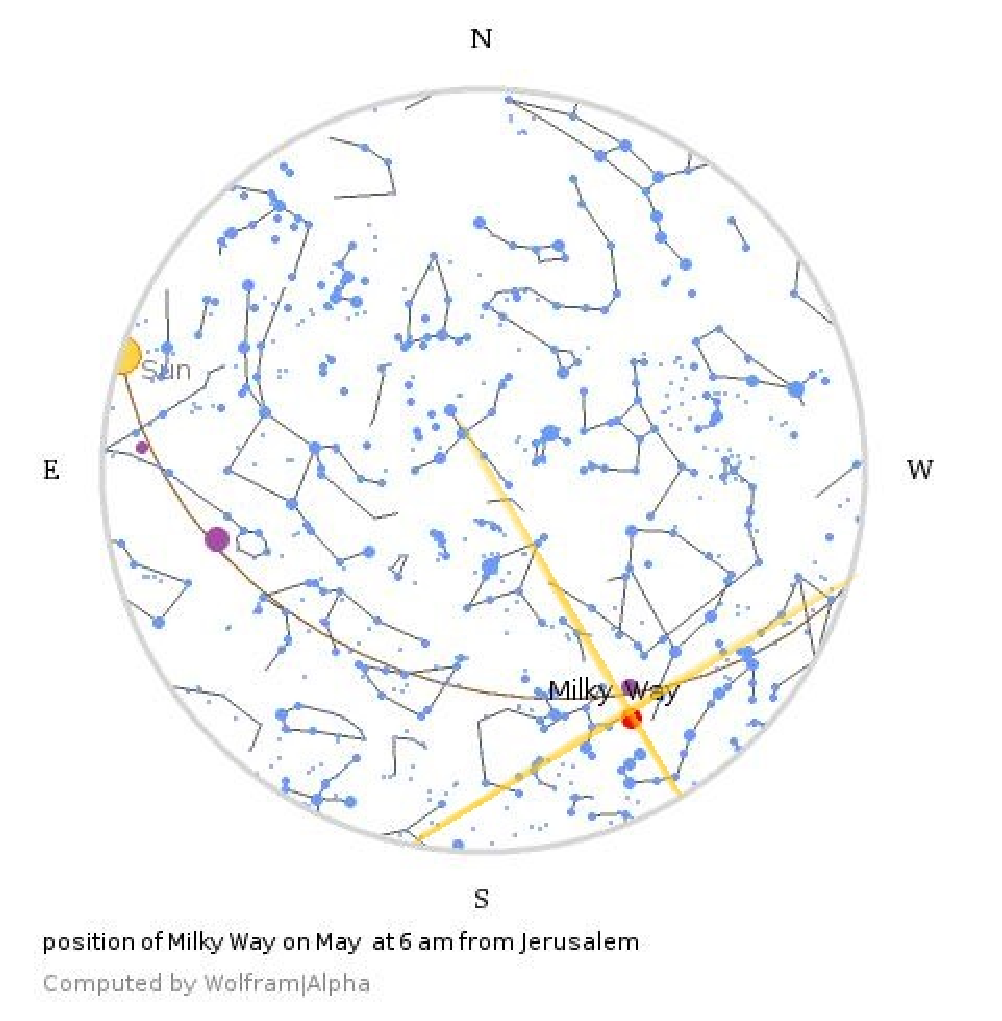}
  \caption{Position of Milky Way on May 7 at 6 am in Jerusalem.}
 \label{jerusalem}
\end{figure}

The day time in Late Antiquity was calculated beginning from sunrise; "the third hour in the morning" corresponds to approximately 5 am. Cyril was most likely to observe "the cross" from his residence in Zion region of Jerusalem, located south from Golgotha. Therefore, one ray of light seemed extending up to Golgotha in the West and another one reached the Mount of Olive on the East of Jerusalem. 



 \subsection{Russian chronicles of the times of mongol-tatar yoke}

About a thousand years later, in 1317, a flash of three rays was attested in Tver'(Moscow region).
The event was registered in Nikonovskaya Letopis'\cite{nikonletopis}, where three rays of lights were seen in the sky. The position of the Milky Way center cannot be observed in this date. Nevertheless, the projection of the lighted Galaxy disk could have been extended up to the Cygnus and a part of jet went perpendicularly to this plane from the East, see figure~\ref{tver}. 

\begin{figure}[htpb]
  \centering
  \includegraphics[scale=0.5]{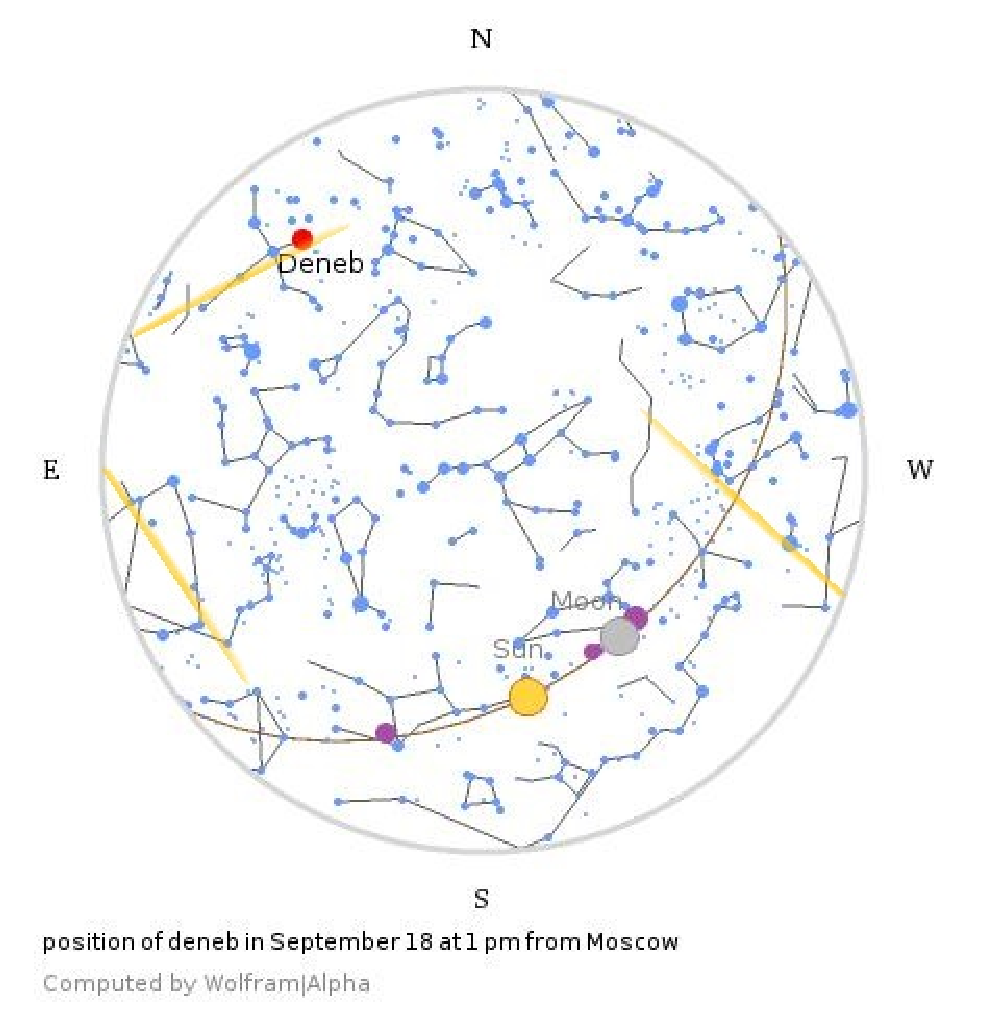}
  \caption{Position of the Milky Way on September 18 from Tver', Moscow proximity.}
  \label{tver}
\end{figure}

As the figure shows, the back side jet, which went around the Earth, appeared from the western side of horizon. That is exactly corresponding to the record of chronicle: "two rays were from East and one ray came from West..."!

Recently, the description of another event was found in the book about Koulikovskaya battle that refers to the original chronicle under 1377 year \cite{rogozhskii}. This time, the crossing rays flared at night. If we account the position of the Moon, which is described in the record, it happened near 2 am on May 5th. The Moon was near the lightening rays, as it is shown on the Wolfram map of the sky, see figure~\ref{moscow1377}, that is exactly corresponding to the description of event.

\begin{figure}[htpb]
  \centering
  \includegraphics[scale=0.6]{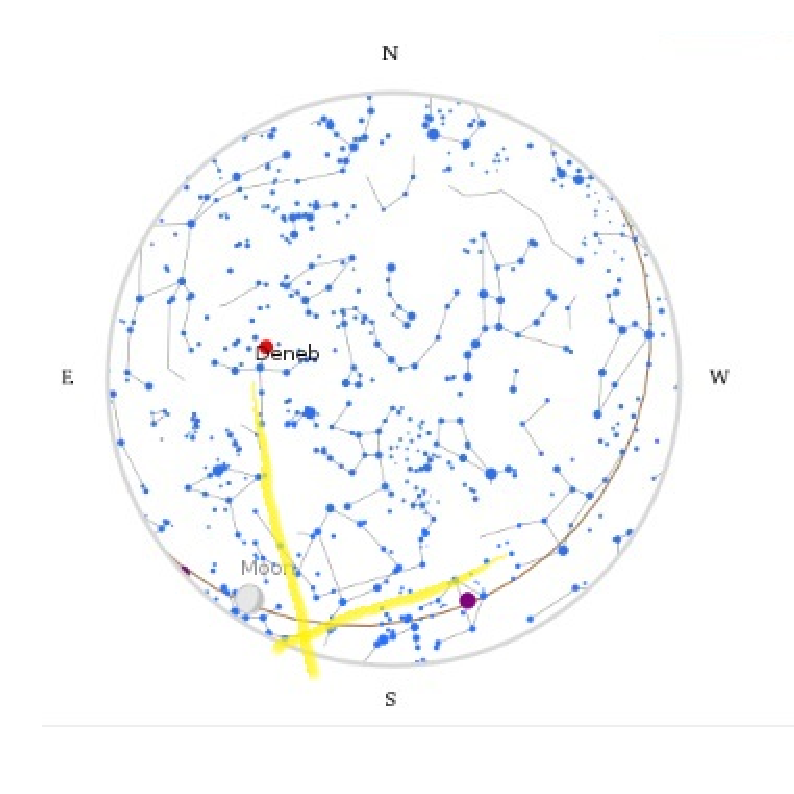}
  \caption{Position of the Milky Way at the night on May 5 of 1377 from Moscow.}
  \label{moscow1377}
\end{figure}


 


\section{More evidences from the ancient authors.}
We could expect that Plato, the great ancient author and scientist, certainly has to write on this outstanding phenomenon. But he could not be an eyewitness along his live. Nevertheless, Plato mentioned about X cross in the heaven in his well-known dialogue "Timaeus" \cite{plato}. On the opinion of G. Latura Beke \cite{latura2009}: "At the end of Timaeus, Plato proclaims that the living Cosmos (which he said had the shape of an X) is a visible, discernible god." 
The vision of lightening cross was mentioned again during 4th century in the time of Emperor Julian Philosopher (361-363 CE). The short mention about this event can be found in \cite{theophanes}: "In these times the holy Cross was seen shining in the heaven from Golgotha to the holy Mount of Olives, circled by the wreath of light; it was even brighter than in the time of Constantius."

\section{Possible astrophysical explanation}

An explanation is possible that the jets went from a neutron star situated on the line of sight to MW center. The developing of such jets doesn't take hundreds of thousands light years, as in SMBHs. The flash of visible light could happen, if neutron star radiation is changing from the radio diapason to roentgen wave lengths and back. But it should not outshine the Sun light.
Another idea came recently after the observation of the roentgen radiation from Sagittarius A* with XMM-Newton experiment \cite{chimneys}. The length of the crossing x-ray "chimneys", which have been observed inside Fermi bubbles. In the past, these crossing flows of x-rays can flash during few hours with the visible light. Our Sun is located on the distance of 26 thousands of light years from MW center, such a way the end of one jet should be seen under almost 45 degrees to the Galaxy disk, taking the quarter of visible sky! The thousand year gap between events is considered as some periodical process at Galaxy center that triggers the x-ray flares. When x-ray sources lost energy, the jets pass into radio diapason after the short time of flashing in visible light. Some recent observations in the arrea that is close to the center of Milky Way shows the traces of radiation which was substationaly more powerful than nowadays. 

\section{Conclusions}

More than four recorded events with the light cross at the day time match with the positions of Galaxy disk on the date and at the place of each vision, \cite{jetactivity}. No other processes in atmosphere can imitate such powerful phenomena, which is as bright as Sun.
We suggest that the SMBH in the center of Milky Way can time-to-time radiate light in visible rays. The "lightening cross" during the day time recorded in 312, 351, 1317 and in 1377 years were caused by the line of two jets beamed back-to-back from the central black hole and crossed the visible projection of the Galaxy disk. What about Moscow events, the possibility to see Milky Way center exists only on May and September.

First time, this event was recorded in the story of power usurpation in Rome by Constantine the Great in October of 312 CE. The sign of "Chi-Rho" on the coins in the times of Constantine is the witness of memorable vision.

The vision in Jerusalem in 351 CE has the description full of details in the letter of Cyril of Jerusalem to Emperor, see in Appendix. The position of X-cross corresponds to the position of MW center early morning on May 7. One jet extends to the West up to Golgotha and MW disk projection lighted the Mount of Olives.
There possibly was another event few years later in the time of Emperor Julian (361-363 CE), but the date and position of X cross was not described well.

The more complicated case happened near Moscow in September 1317. The intersection point cannot be seen at noon, but the part of jet and part of Galaxy disk are seen on the East as two rays of light. At the same time an opposite jet from SMBH went around the back side of Earth and appeared on the West. This event demonstrates that the observations of jet activity are possible from the region of Moscow.

The night event in the Moscow sky was recorded as well in May of 1377 \cite{rogozhskii} and is also well described with the sky maps of Wolfram Astronomy application. The next events are expected in 24th century.

Now, the variety of experimental data on jet activity from central SMBH of Milky Way \cite{recentstudy} gives us the possibility to explain the bright flashes of crossing jets in the past.

\section{Challenges}
 
The future study of the visible jet activity from Milky Way meets the following challenges. 

1. We don't know which astrophysical process caused the thousand year period between events. It may be another BH that is crossing accretion disk under some angle twice in every thousand years. The center of MW is the dense area, which is full of orbital movements that can trigger the "smoking" from chimneys \cite{chimneys}.  

2. The view of events from South America have to be reconstructed, because the jets that flared at night should have considerable impact on the Indian-American civilization in South America and has to be reflected in the ancient religious artefacts. 

3. This knowledge must be retained up to 24th century. 

4. Since the outbursts in the center of MW should also push the particles (or Dark Matter) waves to periphery of the Galaxy, the periodic changes of climate of order few hundreds years can be observed within each millennium.

5. It is also important as well that the 4th and the 14th centuries were times, when climate has become colder that may be correspond to the short periods about 200 years before the flare of jets, while no light or particle radiations should come . In such a way, we have still enough time to study previous historical climate records, where the thousand year periods have to be seen due to the climate variations. Yes, the recent multy 
sciences research \cite{haldon} did show the minimum of temperatures between 310 and 390 years of CE! It is exactly the same period, when the lightening X-cross has been watched in ancient Roman Empire.

\section{Appendix} 

The citations from historical sources would be interesting for more attentive readers. Since it is not reasonable to cite in the ancient languages of original texts, we give the citations in English and in Russian from the contemporary books.

I. The description of Constantine's vision from Eusebius Pamphilius \cite{eusebius}:

"In this state of uncertainty, as he was marching at the head of his troops, a preternatural vision, which transcends all description, appeared to him. In fact, about that part of the day when the sun after posing the meridian begins to decline towards the west, he saw a pillar of light in the heavens, in the form of a cross, on which were inscribed these words, By This Conquer. The appearance of this sign struck the Emperor with amazement and scarcely believing his own eyes, he asked those around him if they beheld the same spectacle; and as they unanimously declared that they did, the Emperors mind was strengthened by this divine and marvelous apparition. On the following night in his slumbers he saw Christ who directed him to prepare a standard according to the pattern of that which had been seen; and to use it against his enemies as an assured trophy of victory. In obedience to this divine oracle, he caused a standard in the form of a cross to be prepared, which is preserved in the palace even to the present time: and proceeding in his measures with greater earnestness, he attacked the enemy and vanquished him before the gates of Rome, near the Mulvian Bridge..." 

II. The episode with "lightening cross" from the letter of Cyril of Jerusalem \cite{cyril}:

"In these holy days ofthe Easter season, on 7 May at about the third hour, a huge cross made of light appeared in the sky above holy Golgotha extending as far as the holy Mount of Olive. It was noy revialed to one or two people alone, but it appeared unmistakably to everyone in the city. It was not as if one might conclude that one have suffered a momentary optical illusion;  it was visible to the human eye above the earth for several hours. The flashes it emitted outshone the rays of the sun, which would have outshone and obscured it themselves if it had not presented the wathers with a more powerful illumination than sun. It prompted the hole populace at ones to run together into the holy church, overcome both the fear and joy at the divine vision. Young and old, men and women of every age, even young girls confined to their rooms at home, natives and foreigners, Christian and pagans visiting from abroad, all todather as if with a single voice raised a hymn of praise to God's Only=begotten Son the wonder-worker. They had the evidence of their own senses that the holy faith of Christians is not based on the persuative arguments of philosophy but on the revelation of Spirit and power (cf. 1 Cor 2.4); it is not proclaimed by mere human beings but testified from heaven by God himself."


 

\smallskip
\end{document}